**Field-free switching of perpendicular magnetic tunnel junction by the interplay of spin orbit and spin transfer torques**

Mengxing Wang[1], Wenlong Cai[1], Daoqian Zhu[1], Jimmy Kan[2], Zhengyang Zhao[3], Kaihua Cao[1], Zhaohao Wang[1], Zilu Wang[1], Youguang Zhang[1], Tianrui Zhang[1], Chando Park[2], Jian-Ping Wang[3], Albert Fert[1,4], Weisheng Zhao[1]*

Spin-orbit torque and spin-transfer torque are leading the pathway to the future of spintronic memories. However, both of the mechanisms are suffering from intrinsic limitations. In particular, an external magnetic field is required for spin-orbit torque to execute deterministic switching in perpendicular magnetic tunnel junctions; the demand for reduced spin-transfer torque switching current to realize ultralow power is still remaining. Thus, a more advanced switching mechanism is urgently needed to move forward spintronics for wide applications. Here, we experimentally demonstrate the field-free switching of three-terminal perpendicular-anisotropy nanopillar devices through the interaction between spin-orbit and spin-transfer torques. The threshold current density of spin-transfer torque switching is reduced to 0.94 MA·cm$^{-2}$, which is a significant decrease compared to that in other current-induced magnetization switching mechanisms. In addition, thanks to the interplay of spin-orbit and spin-transfer torques, lower current-induced switching in the conventional two-terminal perpendicular magnetic tunnel junctions is also achieved.

[1]Fert Beijing Institute, BDBC, and School of Electronic and Information Engineering, Beihang University, Beijing 100191, China. [2]Qualcomm Technologies, Inc. San Diego, California 921121, USA. [3]Department of Electrical and Computer Engineering, University of Minnesota, Minneapolis, Minnesota 55455, USA. [4]Unité Mixte de Physique, CNRS, Thales, Univ. Paris-Sud, University of Paris-Saclay, Palaiseau 91767, France. Correspondence and requests for materials should be addressed to W.Z. (email: weisheng.zhao@buaa.edu.cn)

Magnetization switching by the interaction between spins and charges has greatly brightened the future of spintronic memories.[1–6] This has been evident in the rapid development of spin transfer torque-magnetic random-access memory (STT-MRAM) as a mainstream non-volatile memory technology, in which a spin-polarized current is injected into magnetic tunnel junctions (MTJs) for cell programming.[7–18] However, as cell areas scale down to meet density and power demands, conventional STT-MRAM suffers from serious endurance and reliability issues due to the aging of the ultrathin MgO barrier and read disturbance. The challenge of lowering STT switching current densities to further reduce power consumption is still yet to be met.[19–21] The discovery of spin-orbit torque (SOT) switching in heavy metal/ferromagnetic metal/oxide heterostructures by applying an in-plane charge current to three-terminal devices provides a promising alternative mechanism.[22–28] It shows the potential to enhance the endurance and reliability of MRAM, while improving speed and reducing power consumption.[29–32] Thus, considerable research has been triggered to further elucidate the mechanism of SOT switching, which is currently described as magnetic reversal via two vector components, the damping-like (DL) and field-like (FL) torques.[33,34]

Since the demonstration of perpendicular-anisotropy MgO/CoFeB MTJs (p-MTJs), the switching of perpendicular magnetization by SOT has become of particular interest.[33–38] However, an external magnetic field collinear with the charge current is required to execute deterministic switching of p-MTJs. This intrinsic constraint, combined with the three-terminal device configuration, is limiting the practical application of SOT-MRAM.[26–28,35] Great efforts have been made to eliminate the need for the field assistance, *e.g.*, introducing a lateral structural asymmetry,[39] replacing the heavy metal with antiferromagnetic metal,[40–44] or utilizing a dipole field from an additional in-plane magnetic layer in the stack.[45] In particular, B. Koopmans *et al.* and W. Zhao *et al.* proposed a novel writing scheme to realize field-free perpendicular reversal, *i.e.,* the joint effect of SOT and STT.[46–48] This method has been theoretically studied by evaluating the *Landau–Lifshitz–Gilbert* (LLG) equation, but no experimental result has yet been reported.

Here, we experimentally demonstrate the field-free magnetization switching through the interaction between spin-orbit torques and spin transfer torques by fabricating three-terminal p-MTJ nanopillars on top of Ta bottom electrodes. The correlation between SOT and STT threshold currents is systematically studied, where the threshold current density of STT switching is drastically reduced to 0.94 MA·cm$^{-2}$. Further, due to the interplay of SOT and STT, low current-induced switching without magnetic field is also achieved in the conventional two-terminal p-MTJ devices. For a better understanding, macrospin simulation and finite element simulation were performed to comprehensively analyze the inner mechanism.

**Device fabrication and basic property measurements**

The three-terminal p-MTJ devices we study consist of a circular nanopillar on top of a Ta bottom electrode. Fig. 1a schematically illustrates how the magnetization of p-MTJ free layer can be rotated by the interaction between SOT and STT without any external magnetic field. An in-plane charge current ($I_{SOT}$) passes through the bottom electrode from terminal 1 (T1), giving rise to the SOT acting on the adjacent free layer; the current density $J_{SOT}$ is defined using the cross-sectional area of bottom electrode. Meanwhile, another charge current ($I_{STT}$) is injected into the p-MTJ from terminal 2 (T2), which tunnels the MgO barrier and generates STT by the spin polarization of the reference layer; the current density $J_{STT}$ is defined with the junction area.

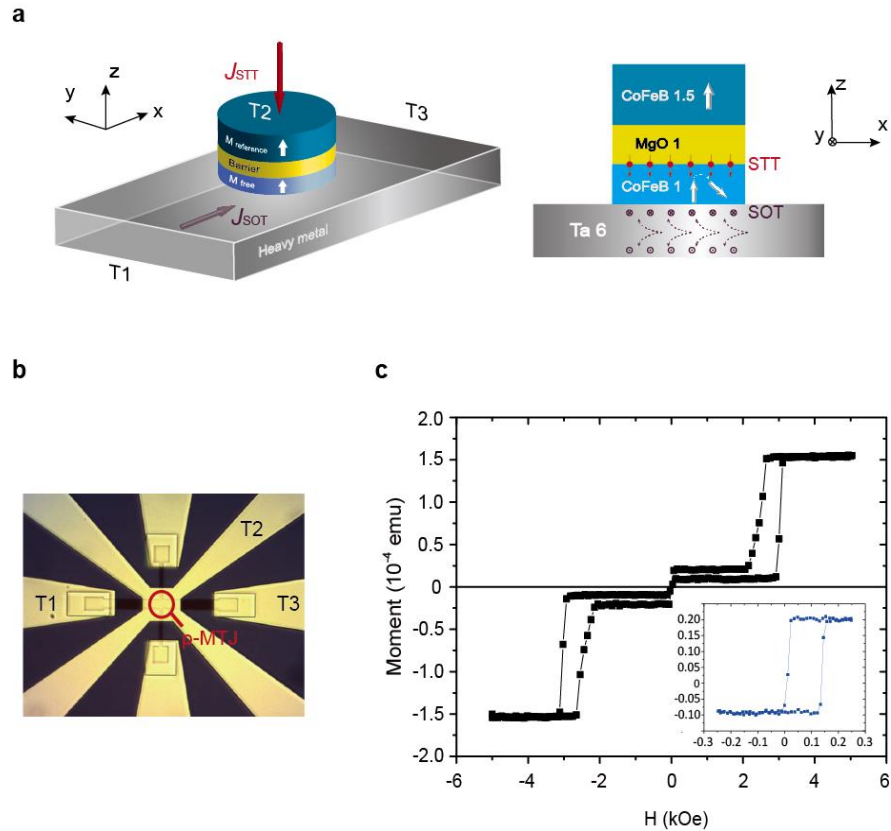

**Figure 1 | Device geometry and Magnetic properties. (a)** Principles of the three-terminal devices. Two currents contribute to the p-MTJ free layer switching: an in-plane current $I_{SOT}$ injected into the bottom electrode from T1 for SOT, and another current $I_{STT}$ input from T2, which is spin-polarized for STT. **(b)** Top view of a three-terminal device by optical microscope. Scale bar indicates 10 μm. **(c)** Out-of-plane hysteresis loops for the blank p-MTJ film after annealing at 325 °C measured by vibrating sample magnetometer (VSM); inset is the minor loop of the free layer.

Our p-MTJ stacks were deposited on thermally oxidized Si substrate by DC/RF magnetron sputtering. The stacks are composed of, from the bottom side, Ta (6.0)/ CoFeB (1.0)/ MgO (1.0)/ CoFeB (1.5)/ [Co/Pt] (5.0)/ Ru (0.85)/ [Co/Pt] (10.2)/ Ru (5.0) (numbers in parenthesis denote layer thickness in nm), and were patterned into three-terminal devices by electron beam lithography and Ar ion milling. As shown in Fig. 1b, the p-MTJ nanopillar with radius ($r$) ranging from 60~200 nm was patterned on top of the Ta bottom electrode with a pair of Hall probes. The radius and the width of the bottom electrode ($w$) are of different values for our investigation. The Hall probes enable the electrical detection of the free layer magnetization direction using the anomalous Hall effect. Those devices were then subjected to 325 °C rapid thermal annealing for 15 min, and present tunnel magnetoresistance ratios around 35% (see Supplementary Note 1 and Supplementary Fig. 1). Fig. 1c shows the M-H hysteresis loops of the blank p-MTJ film under out-of-plane magnetic fields, which indicates the presence of good perpendicular anisotropy in both free layer and reference layer of the p-MTJ stack.

We firstly present the STT switching of the p-MTJ by injecting a pulse current with 1 ms duration ($\tau$) under zero magnetic field, as shown in Fig. 2a. The threshold current density of STT switching ($J_{\text{th-STT}}$) measured here is 5.84 MA·cm$^{-2}$ (also see Supplementary Note 2 and Supplementary Fig. 2). Then, we characterize the SOT switching of our devices. Fig. 2b gives an example of SOT switching by using pulse current with $\tau$ = 200 ms. A constant external magnetic field ($H_x$) of 1 kOe is applied parallel to $I_{\text{SOT}}$ to assist the magnetization reversal. The switching threshold current density of SOT ($J_{th\text{-}SOT}$) in the bottom electrode is measured to be 27.15 MA·cm$^{-2}$. Besides, by performing harmonic Hall voltage measurements towards the Hall bars with identical Ta/CoFeB/MgO structure, the spin Hall angle is estimated at 0.13±0.02 (see Supplementary Note 3 and Supplementary Fig. 3). These results are consistent with the reported experiments regarding similar systems,[26, 37] proving that the CoFeB free layer of our devices can be switched by either $I_{STT}$ or the field assisted $I_{SOT}$.

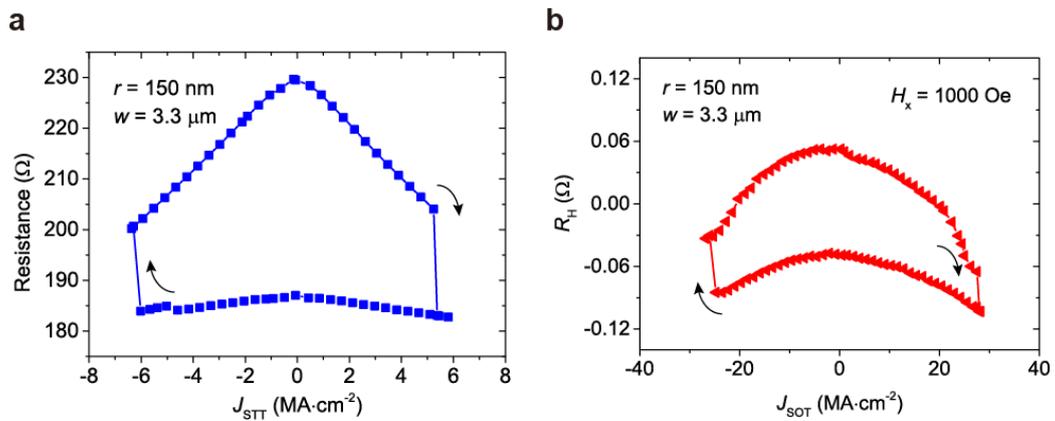

**Figure 2 | Current-induced magnetization switching for a device with $r$ = 150 nm and $w$ = 3.3 μm at room temperature.** (a) STT switching measured by pulse current with $\tau$ = 1 ms. (b) SOT switching with a constant external magnetic field $H_x$ applied along the x axis. Arrows show the perpendicular magnetization transitions from AP to P states or the opposite situation.

**Three-terminal SOT and STT devices**

We investigate the switching behaviour of the three-terminal p-MTJ devices induced by the interplay of STT and SOT. The schematic of the measurement setup is depicted as the inset of Fig.3a, where $J_{SOT}$ and $J_{STT}$ at different ratios are applied simultaneously, and the resistance is recorded to monitor the change of magnetization (also see Supplementary Note 4 and Supplementary Fig. 4). It should be pointed out that no magnetic field is used during the measurement. As shown in Fig. 3a, for a device with $r = 150$ nm and $w = 3.3$ μm, $J_{\text{th-STT}}$ drops dramatically with increasing $J_{\text{SOT}}$ (both with $\tau = 200$ ms); in particular, $J_{\text{th-STT}}$ is reduced to 2.35 MA·cm$^{-2}$ for $J_{\text{SOT}} = 8.3\, J_{\text{STT}}$. The dependence of $J_{\text{th-STT}}$ on $J_{\text{th-SOT}}$ in the case of direct current (DC) measurement is plotted in Fig. 3b. With the assistance of SOT, $J_{\text{th-STT}}$ as low as 0.94 MA·cm$^{-2}$ can be achieved, which is a remarkable decrease compared to other current-induced mechanisms,[35,37,49] hence the ultralow power switching is expected.[48] Moreover, the reduction of charge current passing through the MgO barrier can benefit the lifetime of p-MTJ devices.

Further, we performed pulse current measurement with various durations ranging from 100 μs to 10 ms, where the three-terminal p-MTJ devices have $r = 60$ nm. It can be observed in Fig. 3c that pulse current durations have little influence on the interaction between $J_{\text{th-SOT}}$ and $J_{\text{th-STT}}$. $I_{\text{SOT}}$ in different directions is applied with $I_{\text{STT}}$ (see Supplementary Note 6 and Supplementary Fig. 6), where the data asymmetry cannot be explained by Joule heating alone, because Joule heating only depends on the absolute current value. In addition, our devices with $r = 60$ nm possess a thermal stability factor ($\Delta$) of around 100 (see Supplementary Note 7 and Supplementary Fig. 7), which is sufficient to resist thermal activation. Those experiments suggest that thermal effect is not the major factor for the field-free switching. Fig. 3d shows the relationship between $J_{\text{th-SOT}}$ and $J_{\text{th-STT}}$ with respects to different device sizes. Although the threshold current density is size-dependent, which is mainly the result of redeposition, the field-free switching of p-MTJ by the interplay between SOT and STT can still be evidently verified.

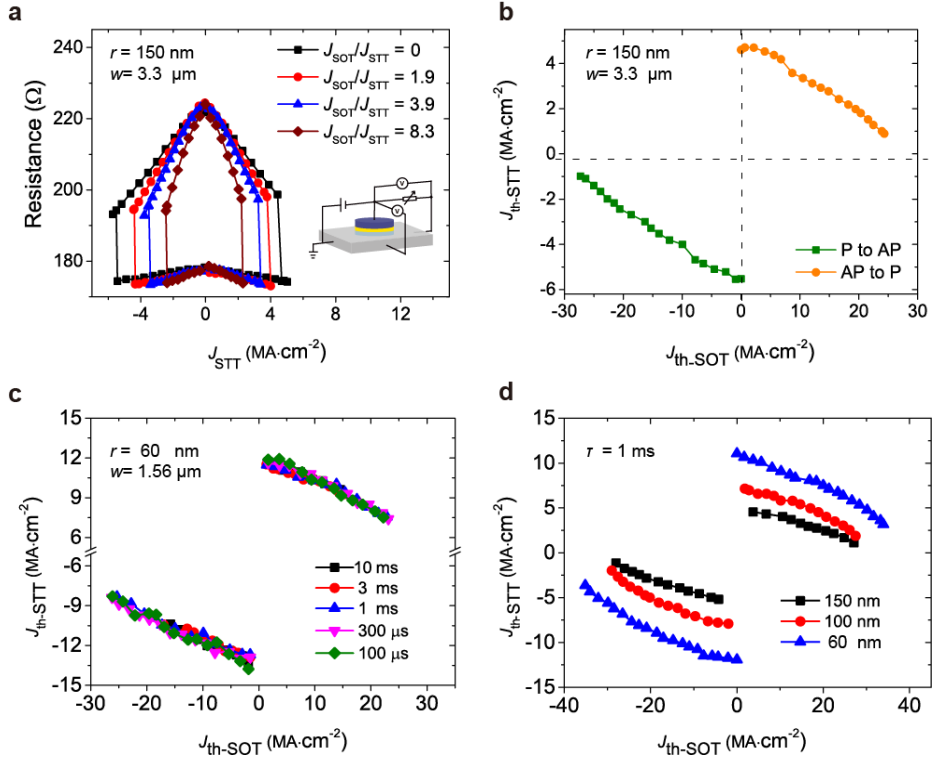

**Figure 3 | Magnetization switching by the interaction between $I_{STT}$ and $I_{SOT}$ for three-terminal devices at room temperature.** **(a)** Magnetoresistances change along with $J_{STT}$ and $J_{SOT}$ at different ratios. Inset is the geometry of our measurement setup, where a resistance box is used to modulate the ratio of two currents. **(b)** $J_{th\text{-}STT}$ measured as a function of $J_{th\text{-}SOT}$ using DC source. **(c)** The relationship between $J_{th\text{-}SOT}$ and $J_{th\text{-}STT}$ measured with various pulse current durations. **(d)** The dependence of $J_{th\text{-}STT}$ on $J_{th\text{-}SOT}$ measured by pulse currents with $\tau = 1$ ms, where the p-MTJ devices have $r = 60, 100,$ and $150$ nm.

**Two-terminal SOT and STT devices**

Interestingly, because of the direct connection between CoFeB free layer and Ta bottom electrode, we speculate that SOT is involved when injecting the $I_{STT}$ alone, *i.e.*, the nominal definition of $J_{SOT} = 0$ in Fig. 3 is actually not so strict. To estimate the impact of SOT in this case, we established two-terminal device measurements with a bipolar external magnetic field $H_x$ along the x axis. The measurement setup is shown in the inset of Fig. 4a. To avoid confusion, we redefine the switching threshold current density going through the p-MTJ in this case as $J_{th}$.

As shown in Fig. 4a, for $H_x > 0$, $J_{th}$ decreases with increasing magnetic field; while for $H_x < 0$, $J_{th}$ presents an opposite trend. This difference indicates the existence of SOT, of which the switching polarity is determined by the directions of both current and field.[25,26] Fig. 4b shows the dependence of $J_{th}$ on various bottom electrode widths. For varying values of *w*, the local current density near the Ta/CoFeB interface changes a lot for a certain current, but the current density within nanopillar remains unchanged. The narrower the bottom electrode is, the smaller the $J_{th}$ becomes, because the contribution from SOT is enlarged.  It is worth noting that according to the fitting curve $J_{th}$ as low as 2.76 MA·cm$^{-2}$ is achieved in the case of *w = 2r*. These results suggest the presence of strong SOT in our two-terminal devices.

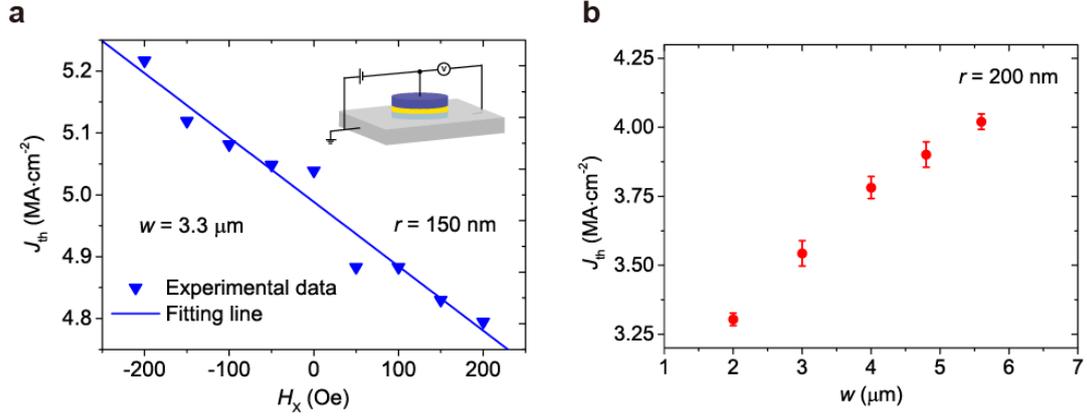

**Figure 4 | Two-terminal device measurements at room temperature. (a)** $J_{th}$ measured from a two-terminal device with $r$ = 150 nm and $w$ = 3.3 μm as a function of $H_x$. Inset is the geometry of our measurement setup. **(b)** $J_{th}$ changes with varying bottom electrode width, while the radius of p-MTJs is fixed as 200 nm. $J_{th}$ here is the average absolute values for parallel-to-antiparallel and antiparallel-to-parallel reversals. The error bars reflect the corresponding uncertainties for the averaging.

**Discussion of physical mechanisms**

The benefit of the joint effect between SOT and STT can be summarized as below: first, the existence of STT, no matter how small, can break the symmetry of SOT, hence the field-free switching of p-MTJ devices is enabled; second, since SOT is orthogonal to the free layer magnetic moment at equilibrium states, the stable magnetic states of free layer will be instantly disturbed once the SOT current is applied. Consequently, the efficiency of STT can be significantly promoted, which means the switching time and power consumption can be reduced.

To fully realize the potential of the scheme, $J_{SOT}$ is supposed to be larger than a critical value for the rapid rotation of the free layer magnetization in equilibrium states, even to the in-plane orientation. With only the DL torque of SOT considered, namely, the driving force for switching only comes from spin Hall effect (SHE), the critical value $J_{C0\text{-SHE}}$ can be approximately derived from a macro model as $J_{C0-SHE} = \frac{e\mu_0 M_S H_{K,eff} t_F}{\hbar \theta_{SH}} = \frac{2eK_B T}{\hbar \theta_{SH} \pi r^2} \cdot \Delta$,[29] where $e$ denotes the elementary charge, $\mu_0$ the permeability of vacuum, $\hbar$ the reduced Plank constant, $\theta_{SH}$ the spin Hall angle, $K_B$ the Boltzmann constant, $T$ the environment temperature, $M_S$ the saturation magnetization, $H_{K,eff}$ the effective perpendicular anisotropy field, and $t_F$ the thickness of the free layer, respectively. Note that thermal stability factor $\Delta = \frac{\mu_0 M_S H_{K,eff} t_F \pi r^2}{2K_B T}$ of more than 60 is required to satisfy the industry standard retention time of 10 years. Fig. 5a shows $J_{C0\text{-SHE}}$ and $H_{K,eff}$ as a function of $r$ when $\Delta=60$ is imposed at room temperature. It is obvious that $J_{C0\text{-SHE}}$ and $H_{K,eff}$ will dramatically increase as the size of a p-MTJ shrinks. Particularly, for $r < 20$ nm, $H_{K,eff}$ ought to be higher than 0.35 T and $J_{C0\text{-SHE}}$ exceeds 450 MA·cm$^{-2}$, which is unrealistically high. Therefore, the magnetization switching induced by SHE assisted STT or by the SHE alone seems impractical for p-MTJ with large effective anisotropy, which is exactly the case for our devices with a measured $\mu_0 H_{K,eff}$ of more than 400 mT.

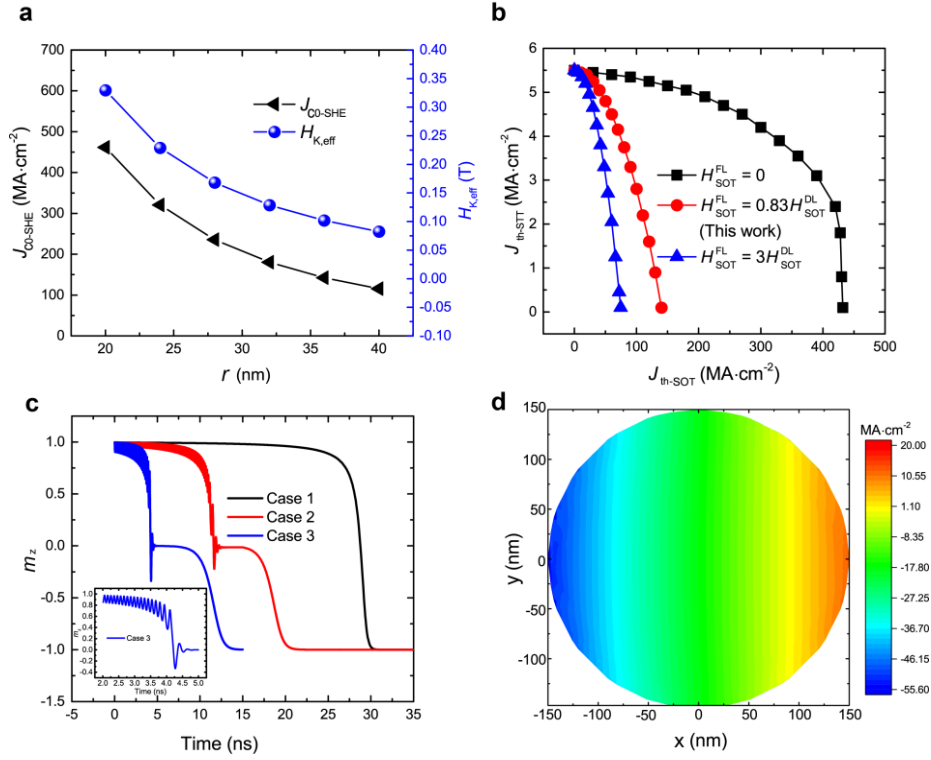

**Figure 5 | Results of macrospin simulation and finite element simulation. (a)** Dependence of $J_{C0\text{-SHE}}$ and $H_{K,\text{eff}}$ on $r$ when $\Delta = 60$ is imposed at room temperature. **(b)** Relationships between $J_{\text{th-STT}}$ and $J_{\text{th-SOT}}$ under different ratio of $H_{\text{SOT}}^{\text{FL}}$ to $H_{\text{SOT}}^{\text{DL}}$, and the ratio is estimated at 0.83 in this work. **(c)** Time evolution of $m_z$ under varying SOT and STT current density when $\frac{H_{\text{SOT}}^{\text{FL}}}{H_{\text{SOT}}^{\text{DL}}} = 0.83$. Case 1: $J_{\text{SOT}} = 0$, $J_{\text{STT}} = 5.7$ MA·cm$^{-2}$; Case 2: $J_{\text{SOT}} = 80$ MA·cm$^{-2}$, $J_{\text{STT}} = 4.15$ MA·cm$^{-2}$; Case 3: $J_{\text{SOT}} = 150$ MA·cm$^{-2}$, $J_{\text{STT}} = 0.1$ MA·cm$^{-2}$. The insert shows the detailed reversal behaviour for Case 3 from 2 ns to 5 ns. **(d)** Detailed current projection along the x axis at the interface between p-MTJ and bottom electrode for $J_{\text{STT}} \sim 3$ MA/cm$^2$ and $J_{\text{SOT}} \sim 16$ MA/cm$^2$, where the current goes in -x direction, and the p-MTJ is set as $r = 150$ nm. (For parameters used in simulations, please see Supplementary Note 8)

However, the FL torque of SOT should not be neglected since recent studies have shown its crucial role in promoting the domain wall motion and magnetization reversal process, especially in Ta/CoFeB/MgO systems where large FL torque exists. Therefore, to accurately clarify the interplay of SOT and STT in our devices, the FL torque of SOT is also involved. Within macrospin approximation, the following LLG equation with additional SOT and STT terms is numerically solved to depict the free layer dynamics:

$$\frac{\partial \boldsymbol{m}}{\partial t} = -\gamma\mu_0(\boldsymbol{m}\times \boldsymbol{H}_{\text{K, eff}}) + \alpha\left(\boldsymbol{m}\times\frac{\partial \boldsymbol{m}}{\partial t}\right) + \gamma\mu_0 H_{\text{STT}}^{\text{DL}}(\boldsymbol{m}\times \boldsymbol{m}_{\text{p}}\times \boldsymbol{m}) + \gamma\mu_0 H_{\text{STT}}^{\text{FL}}(\boldsymbol{m}\times \boldsymbol{m}_p)$$
$$+ \gamma\mu_0 H_{\text{SOT}}^{\text{DL}}(\boldsymbol{m}\times \boldsymbol{\sigma}\times \boldsymbol{m}) + \gamma\mu_0 H_{\text{SOT}}^{\text{FL}}(\boldsymbol{m}\times \boldsymbol{\sigma})$$

where $\boldsymbol{H}_{\text{K, eff}} = H_{\text{K, eff}} m_z \hat{z}$ denotes the effective anisotropy field, including the anisotropy field and demagnetization field, $\boldsymbol{m} = \frac{\boldsymbol{M}}{|M_s|}$ the reduced magnetization, $\gamma$ the gyromagnetic ratio, $\alpha$ the Gilbert damping factor, $\boldsymbol{m}_P$ and $\boldsymbol{\sigma}$ the electron polarization direction of the spin polarized current arising from STT and SOT terms, respectively. Both DL and FL torque terms are considered with $H_{\text{STT}}^{\text{DL}}$, $H_{\text{STT}}^{\text{FL}}$, $H_{\text{SOT}}^{\text{DL}}$ and $H_{\text{SOT}}^{\text{FL}}$ representing the corresponding current-dependent proportionality constants. Fig. 5b shows the dependence of $J_{\text{th-STT}}$ on $J_{\text{th-SOT}}$ under different ratio of $H_{\text{SOT}}^{\text{FL}}$ to $H_{\text{SOT}}^{\text{DL}}$, from which we can see that the $J_{\text{th-SOT}}$ required to minimize $J_{\text{th-STT}}$ depends largely on the ratio of FL torque to DL torque of SOT. For instance, when $\frac{H_{\text{SOT}}^{\text{FL}}}{H_{\text{SOT}}^{\text{DL}}} = 0.83$ (see Supplementary Note 3 and Supplementary Fig. 3), as observed in our experiments, $J_{\text{SOT}}$ ~140 MA·cm$^{-2}$ is needed, which is a three-fold reduction compared to the case where FL torque is excluded. Furthermore, $\frac{H_{\text{SOT}}^{\text{FL}}}{H_{\text{SOT}}^{\text{DL}}} > 3$ has been recently reported, and in that case $J_{\text{SOT}}$ calculated from macrospin simulation is lower than 80 MA·cm$^{-2}$. Under this circumstance, it is a feasible approach to use the interplay of SOT and STT to realize both field-free switching and low STT current density.

Fig. 5c displays the time evolution of the reduced perpendicular component of free layer magnetization $m_z$ under varying SOT and STT current density when $\frac{H_{\text{SOT}}^{\text{FL}}}{H_{\text{SOT}}^{\text{DL}}} = 0.83$. As $\frac{J_{\text{SOT}}}{J_{\text{STT}}}$ gradually increases, the retention time of the reversal process decreases to a large degree. Besides, if $J_{\text{SOT}}$ is large enough, the switching behavior is SOT dominant, and the magnetization will stay in the nearly in-plane orientation with $m_z$ being slightly negative. After the current is turned off, deterministic switching can be realized due to the damping of magnetization precession towards –z direction. From the inset of Fig. 5c, we can also see the impact of FL torque of SOT on the switching behavior. Since the effective field of this term is along the y axis, this term tends to rotate the magnetization ***m*** around the y axis, thus making it easier for ***m*** to cross over the in-plane orientation. In contrast, the FL torque of STT does not contribute significantly to the p-MTJ switching, because the effective field of the torque is along the z axis, and the magnitude of $H_{\text{STT}}^{\text{FL}}$ is negligible compared to the large $H_{\text{K,eff}}$ of our devices.

Note that the $J_{\text{SOT}}$ utilized in our experiments to reduce the $J_{\text{STT}}$ is lower than that calculated from the macro model, especially for the p-MTJ devices with larger size. We attribute this phenomenon to the following reasons. Firstly, macro models do not accurately apply to the p-MTJs with r > 60 nm as the whole switching process comprises domain wall nucleation at the edges and propagation through the free layer, owing to which the current density required is somewhat reduced. Secondly, the current distribution within the devices is nonuniform in the experiments but homogeneous in the macro model simulations. Fig. 5d estimates the current distribution in the experiments, showing a detailed current projection along the x axis at the interface between p-MTJ and bottom electrode for $J_{STT}$ ~3 MA·cm$^{-2}$ and $J_{SOT}$ ~16 MA·cm$^{-2}$. It is obvious that the local current density distributes with gradient and the maximum value is estimated to be ~60 MA·cm$^{-2}$, which is much higher than that in the uniform part of the Ta electrode.

To conclude, we experimentally realized the deterministic field-free magnetization switching induced by the interplay of spin-orbit torques and spin transfer torques in three-terminal devices with p-MTJ nanopillars and Ta bottom electrodes. By the systematic profile of the relationship between $J_{STT}$ and $J_{SOT}$, a $J_{th-STT}$ as low as 0.94 MA·cm$^{-2}$ is achieved, which can benefit significantly the power consumption and lifetime of p-MTJ nanopillars. The interaction is also observed in conventional two-terminal p-MTJ devices, which provides a potential pathway towards high-density MRAM. Additionally, macrospin simulations and finite element calculations were performed to give a clear view of the inner mechanism.

**Methods**

**Device fabrication and measurements.** Our p-MTJ films were grown on thermally oxidized Si substrate by DC magnetron sputtering, while MgO deposition was performed by RF sputtering. For the three-terminal device patterning, nanopillars were defined by Vistec EBPG5000+ e-beam lithography at the centre of Ta bottom electrodes with various widths followed by Ar ion milling. Then they were fully covered with SiO$_2$ for insulation. After the lift-off procedure, via holes were made over the bottom electrodes. Both the bottom electrodes and p-MTJs were then connected to 90 nm Ti/Au electrodes to allow electrical contact for measurement using e-beam evaporation. The blank stacks were studied with VSM and ferromagnetic resonance systems. The setup for current-induced p-MTJ switching consists of a Lake Shore CRX-VF cryogenic probe station, Keithley 6221 current sources, and 2182 nanovolt meters.

**Data availability.** The data that support the plots within the paper and other findings of the study are available from the corresponding authors upon reasonable request.

**Code availability.** The code used for macrospin simulations of this study are available from the corresponding authors upon reasonable request.

**Acknowledgements**

The authors gratefully acknowledge the National Natural Science Foundation of China (Grant No. 61571023, 61627813), the International Collaboration Project B16001, and the National Key Technology Program of China 2017ZX01032101 for their financial support of this work. The device fabrication that was carried out at the University of Minnesota Nanofabrication Centre, which receives partial support from NSF through the NNIN program. J.-P.W. thanks the Robert F. Hartmann Endowed Chair Professorship.

**Author contributions**

W.Z. initialized, conceived and supervised the project. M.W., W. C. and D.Z. contributed equally to this work. M.W. and Z.Z fabricated the devices under J.-P.W.'s supervision. W.C. and M.W. and T.Z. performed the measurements. J.K. and C.P developed, grew and optimized the films. D.Z. performed the macrospin and finite element calculations, and A.F. analysed the results. M.W., W.C., D.Z. and W.Z. wrote the manuscript. All authors discussed the results and implications.

**Competing financial interests:** The authors declare no competing financial interests.